\documentclass[11pt,a4paper]{article}
\pdfoutput=1
\usepackage[hyperref]{emnlp2020}
\renewcommand{\UrlFont}{\ttfamily\scriptsize}
\usepackage{latexsym}
\usepackage{times}  
\usepackage{bbm}
\usepackage{amsthm}
\usepackage{amsmath,amssymb}
\usepackage{mathtools}
\usepackage{color}
\usepackage{colortbl}
\usepackage{graphicx}
\usepackage{booktabs}
\usepackage{xcolor,colortbl}
\usepackage{paralist}
\usepackage{xspace}
\definecolor{mygray}{gray}{0.85}
\usepackage{pgfplotstable}
\newcommand{\bilstm}{\textsc{bilstm}\xspace}
\newcommand{\lstm}{\textsc{lstm}\xspace}
\newcommand{\gru}{\textsc{gru}\xspace}
\newcommand*{\affaddr}[1]{#1} 
\newcommand*{\affmark}[1][*]{\textsuperscript{#1}}
\newcommand*{\email}[1]{\texttt{#1}}
\pgfplotsset{compat= 1.13}
\usepackage{adjustbox}
\usepackage{url}
\usepackage{microtype}
\aclfinalcopy 
\def\aclpaperid{1322} 

\title{NwQM: A neural quality assessment framework for Wikipedia}

\author{%
  Bhanu Prakash Reddy\thanks{*Authors contributed equally} \affmark[1],Sasi Bhushan\affmark[*]\affmark[2],Soumya Sarkar\affmark[*]\affmark[3],Animesh Mukherjee\affmark[4] \\ 
\affaddr{IIT Kharagpur, India\affmark[2,3,4], Adobe Research, India\affmark[1]}\\
\email{soumya015@iitkgp.ac.in\affmark[3],guda@adobe.com\affmark[1]}\\
\email{sasibhushan3@gmail.com\affmark[2], animesh@cse.iitkgp.ac.in\affmark[4]}
}

\date{}

\begin{document}
\maketitle
\begin{abstract}
 Millions of people irrespective of socio-economic and demographic backgrounds, depend on Wikipedia articles everyday for keeping themselves informed regarding popular as well as obscure topics. Articles have been categorized by editors into several quality classes, which indicate their reliability as encyclopedic content. This manual designation is an onerous task because it necessitates profound knowledge about encyclopedic language, as well navigating circuitous set of {\em wiki} guidelines. In this paper we propose {\bf N}eural {\bf {w}}ikipedia {\bf Q}uality {\bf M}onitor (NwQM), a novel deep learning model which accumulates signals from several key information sources such as \textit{article text}, \textit{meta data} and \textit{images} to obtain improved Wikipedia article representation. We present comparison of our approach against a plethora of available solutions and show {$8\%$} improvement over state-of-the-art approaches with detailed ablation studies.     
\end{abstract}

\section{Introduction}
Wikipedia is one of the most prominent sources of free information in the world today. Since reliable and advertisement free material is mostly behind pay-walled sources, a huge volume of global population is directed toward this extraordinary crowd sourced platform for information ranging from history, politics, pop-culture to even scientific topics~\cite{horta2020sudden}. 

Although  Wikipedia has grown significantly in terms of volume and veracity over the last decade, the quality of articles is not uniform~\cite{warncke2015success}. The quality of Wikipedia articles is monitored through a rating system where each article is assigned one of several class indicators. Some of the {\em major }article categories are \textbf{FA}, \textbf{GA}, \textbf{B}, \textbf{C}, \textbf{Start} and \textbf{Stub}. Most complete and dependable content is annotated by an FA ({\em aka featured article}) tag while lowest quality content is annotated with a Stub tag. The intention behind this elaborate scheme is to notify editors regarding current state of the article and extent of effort needed for escalating to encyclopedic standards\footnote{\UrlFont{wiki/Wikipedia:WikiProject Wikipedia/Assessment}}.

There exist several guidelines which direct editors in annotating articles into respective classes. Some of the traits of a FA article are engaging and comprehensive prose with neutral point of view and verifiable claims. It must also rigorously follow the style manual, i.e. the page structure. Further the content should be stable, i.e. devoid of edit warring. Understanding compliance with these guidelines often require detailed knowledge about language usage as well as domain knowledge about Wikipedia page layout and style principles. Often it is nontrivial to discern qualifying differences between articles which merits their ratings without inculcating personal biases. 

Consider the wikipage of two prominent US presidents \textit{Abraham Lincoln} (GA) and \textit{John F. Kennedy} (B). Both pages are indistinguishable in-terms of coverage and engagement, however on closer assessment it is apparent that President Kennedy's page contains unattributed opinion such as the statement {\em This crisis brought the world closer to nuclear war than at any point before or after ...} in the {\em Cuban Missile Crisis} section. It also has vague quantifiers such as {\em some questioned, some crtics, somewhat successful etc.} Similarly, if we look at wikipages of historical figures \textit{Akhenaten} (B) and \textit{Cleopatra} (FA) it is difficult to discern their quality just from the content. A deep dive into the individual \textit{talk pages} reveal that the former page has unresolved content issues and disputes which justifies the given rating. Hence it is a difficult task to manually judge language specific nuances present in the main page text, topic level disputes manifesting in the talk pages as well as section layout and image positioning before deciding a correct rating. This is also apparent from the quality statistics\footnote{\UrlFont{wiki/Wikipedia:Good article statistics}} which shows that only $0.09\%, 0.5\%$ of the $\sim 6M$ English Wikipedia articles have a FA and a GA tag respectively.

Current approaches for automatic quality assessment by {\em Wikimedia foundation\footnote{\UrlFont{ wikimediafoundation.org}}} use handcrafted features from main article text for classifying  quality class~\cite{halfaker2019ores}. Analogous approaches exist~\cite{dang2016quality} which attempt to automatically generate features from main text using deep learning models such as {\em doc2vec}~\cite{le2014distributed}. Other approaches use deep sequence models such as \bilstm~\cite{shen2017hybrid} as well as combining representations obtained with additional modalities like image~\cite{shen2019joint}. The principal focus of existing works have been concentrated on main article text. However one of the key sources of metadata about an article, i.e., the corresponding \textit{talk page} has been ignored. Talk pages contain crucial information concerning stability of a page. They also hold evidence whether discussion threads encompassing topics are decisive. Besides, representation of main page text using sequence models cannot capture high level semantic signals such as whether the introduction section is a summary, whether the coverage of topics is polarized or the information is redundant or the wording is convoluted.

In this paper we propose {\bf N}eural {\bf {w}}ikipedia {\bf Q}uality {\bf M}onitor (\textbf{NwQM}) which integrates information from multiple sources, i.e., main page text, metadata and html rendering resulting in improved quality assessment. We use bidirectional contextual representation~\cite{devlin2018bert} for encoding article text. However, contrary to document representation using BERT~\cite{adhikari2019docbert}, which is not adequate for large text documents, we first segment articles organically based on sections. We fine tune on each section text individually followed by a summarization layer which preserves the sequential nature of sections. Similar representation of atomic units and summarisation is applied on talk pages. We also obtain images from the raw markup using Imagekit\footnote{\UrlFont{pypi.org/project/imgkit}}. These images are further embedded in a vector space using Inception V3~\cite{szegedy2016rethinking}. Inception V3 is pre-trained on Imagenet~\footnote{\UrlFont{www.image-net.org/}} and we fine tune on our dataset to cater to our task. Our experiments show that combining information sources from these sources leads to improved result eclipsing current state of the art ~\cite{shen2019joint} by $8\%$.

Our main contributions are enumerated below.
\begin{compactenum}
    \item We propose a multimodal framework from quality assessment of Wikipedia articles which leverages contextual representation obtained from bidirectional transformers and supports conditional summarization.
    \item To the best of our knowledge this is the first work which utilizes meta pages, i.e., talk pages as an additional signal for this task. All code, sample data and image embeddings related to the paper are made available\footnote{\url{https://github.com/sasibhushan3/NwQM\_EMNLP}}$^\text{,}$\footnote{\url{https://zenodo.org/record/4066405}} to promote reproducible research.
\end{compactenum}

\section{Related work}
Automatic article assessment is one of the key research agendas of the Wikimedia foundation\footnote{\UrlFont{www.mediawiki.org/wiki/ORES}}. One of the preliminary approaches~\cite{halfaker2015artificial} seeking to solve this problem extracted structural features such as presence of infobox, references, level 2 headings etc. as indicators of the article quality. Other approaches explored distributional representation as well as sequence models~\cite{dang2016quality,shen2017hybrid,shen2019joint}. \cite{zhang2018history} attempted to solve this problem by formulating features capturing dynamic nature of the articles. A complementary direction of exploration has been put forward by~\cite{li2015automatically,de2015measuring} where correlation between article quality and structural properties of co-editor network and editor-article network has been exploited.

This task can also be solved by exploring the rich literature of document classification. One of the characteristics of Wikipedia articles are that these are long documents hence sequence models can suffer from~\cite{atkinson2018charlottesville} catastrophic forgetting.~\cite{yang2016hierarchical} proposed to solve this by leveraging a hierarchical organization of documents. Further improvements have been demonstrated by employing bidirectional transformers~\cite{adhikari2019docbert,ostendorff2019enriching}. However to the best of our knowledge no previous work have used metadata about article pages as source of additional signals. Also we investigate information fusion from multimodal sources for generating improved article representation. 
\section {Dataset}
\label{sec:dataset}
\begin{table}[t]\setlength{\tabcolsep}{14pt}
    \centering
    \begin{tabular}{c c}
    \toprule
    Class &  Article count  \\
    \midrule
     FA & 3589  \\
     GA & 5900  \\
     B & 5900  \\
     C & 5900  \\
    Start & 5900  \\
    Stub & 5900  \\
    \hline
    Total & 33089 \\
    \bottomrule
\end{tabular}
		\caption{\label{table_1} Wikipedia dataset of articles with respective {\em talk }pages.}
	\end{table}

Wikimedia foundation stores all data for its multilingual wikiprojects in the form of Wikidumps\footnote{\UrlFont{https://dumps.wikimedia.org}}. We downloaded first $100$ English Wikipedia dumps which are 7z archived {\em xml} files. The combined uncompressed size of these files is $\sim 8TB$ and it contains a sample of $\sim 6M$ English Wikipedia text from the first version to last update as of June 2019. Because of limitation of space we did not go though the entire English Wikipedia archive. Each uncompressed file is approximately of $80GB$ size and have random samples of  approximately $\sim 5k$ Wikipedia pages. We parsed each file using mediawiki xml parser\footnote{\UrlFont{https://pypi.org/project/mwxml}} and in one linear scan we tried to locate if the main page text and talk page text is in the same dump xml file. More specifically if we encounter the main Wikipedia article {\em Cleopatra}, we remember it in a dictionary and seek to locate {\em Talk:Cleopatra} in future scans or vice versa. However it is entirely possible that {\em Talk:Cleopatra} is not present at all in the currently encountered xml file and may be present somewhere else; in such case we ignore that article. If we can locate both main and talk pages of the same article we save it for reference. We include in our corpus maximum number of articles for \textbf{GA}, \textbf{B}, \textbf{C}, \textbf{Start} and \textbf{Stub} while maintaining equality. For \textbf{FA} articles, we included entire extracted corpus, because such articles are scarce.   

Although this process is na\"ive, yet we manage to extract moderately balanced number of datapoints for each class. We would further like to note that few other public datasets exits for this task. The first version is made available by the Wikimedia foundation\footnote{\UrlFont{analytics.wikimedia.org/published/datasets}} which has $30K$ datapoints with approximately $5k$ pages in each of the 6 classes.~\cite{shen2017hybrid} has pointed out that this dataset contains many noisy datapoints, e.g., empty pages labeled as FA class. Other datasets are made available by~\cite{shen2019modelling,shen2019feature,warncke2015success}. Our investigation shows that none of the former datasets contain meta pages which prompts us to extract this novel data ourselves. Also some of the datasets have uniform article length across classes, which is often not the case in reality. For example {\em Dodo} and {\em Grey-necked wood rail
} are both FA articles with very different article length. Considering uniform article length may lead to overfitting. We used stratified random sample of $80\%$ data for training and $10\%$ each for validation and testing. The overall distribution of the articles with respective classes in our dataset is enumerated in Table~\ref{table_1}.

\section{Proposed solution}
In this section we present a detailed description of our multimodal approach. We start this section by explaining the various notations that we use in the subsequent sections of the paper. We then proceed to first describe representation mechanisms of explicit signals about article quality, obtained from main text, i.e., presence of bias, claim verifiability, coherent wording, citations etc. We further elaborate on the mechanisms employed, to capture implicit quality indicators such as article stability, collaborative nature of editors obtained from article talk text as well as visual renderings of documents to capture how well the article follows the style manual. Further, we describe the different ways in which we combine the information to predict the quality of the page. We present the overall architecture of our model in Figure~\ref{fig:overall_archi}.

\subsection{Preprocessing}\label{Sec:preprop}
We first pre-process the Wikipedia articles and talk pages to convert the text from Wiki Markup Language\footnote{\UrlFont{en.wikipedia.org/wiki/Wikipedia:Wiki\_Markup\_Language}} to plain English text format using a python text crawler\footnote{\UrlFont{github.com/attardi/wikiextractor}}. We then replace the meta content in the page such as infobox, level 1 section headings, level 2 section heading, internal wikilink, external link, inline reference, footnote template, image template, quotation template and categories into special tokens which act as additional features using mediawiki parser\footnote{\UrlFont{mwparserfromhell.readthedocs.io}}. We use this pre-processed text in the subsequent models.

\subsection{NwQM overview}
 Our approach is inspired by the hierarchical document representation approach proposed in~\cite{yang2016hierarchical}, designed to capture signals from multiple levels of document organization. Wikipedia pages have an organic structure, i.e., words form sentences, sentences form paragraphs, paragraphs form sections and sections form a page. We build page representation by first generating section embeddings, followed by a suitable summarization. We use \textsc{BERT}~\cite{devlin2018bert} for section representation and bidirectional \gru with self attention for summarization. Explicitly our encoder has two levels in the hierarchy, however implicitly by employing bidirectional transformers~\cite{vaswani2017attention} with special tokens, we can aggregate contextual features across multiple levels.

\begin{figure*}[!ht]
    \centering
    \includegraphics[width=0.93\textwidth]{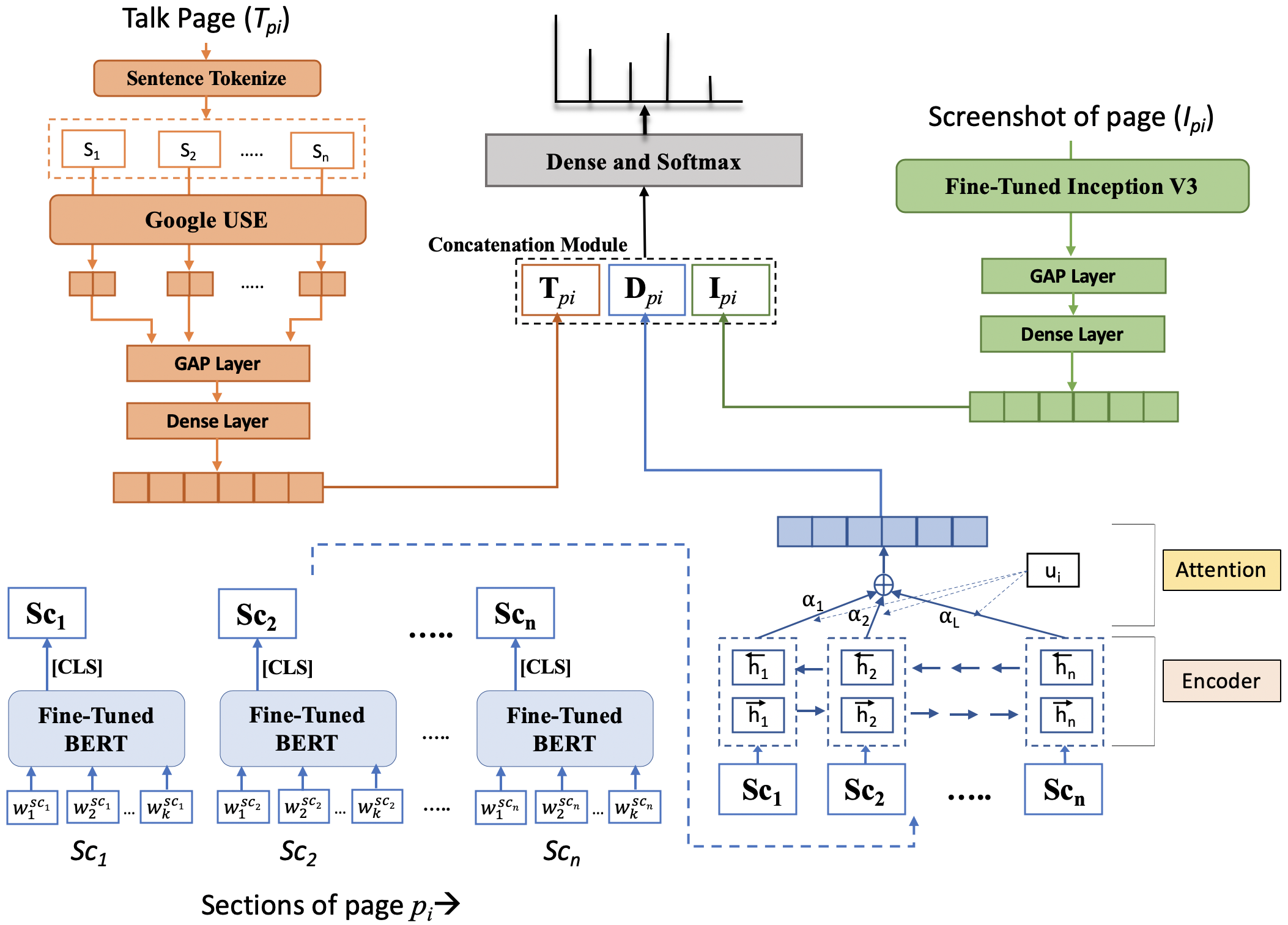}
    \caption{The overall pipeline of \textbf{NwQM}. We have taken cutoff for the number of tokens in sequence i.e $k=512$ and no. of sequence $n=16$. Parameters of fine-tuned \textsc{bert}, Inception V3 are fixed.}
    \label{fig:overall_archi}
\end{figure*}

\subsection{Fine tuning BERT}
We fine tune the BERT~\cite{devlin2018bert} model on the extracted Wikipedia pages for classifying them into the 6 quality categories. We pass the pre-processed textual content of the Wikipedia page to the BERT model and pass the \textsc{[CLS]} token’s representation to a dense followed by softmax layers to classify the page. We follow the fine-tuning strategy proposed by~\cite{sun2019fine} for long input sequences to overcome the limitations of BERT in handling inputs more than 512 tokens. We concatenate the first 128 and the last 384 tokens of the page. For the statistical observations on advantage of this approach please refer to~\cite{sun2019fine}. We use the BERT's tokenizer with additional meta content tokens explained in Section~\ref{Sec:preprop}. We fine tune the model end-to-end (110M parameters). Once the model is finetuned we freeze the weights, thus in subsequent steps of the training the parameters are not updated. 

\subsection{BERT section encoder}
We use previously obtained fine tuned BERT for generating section representations. It has been shown that fine tuned BERT performs extremely well in subject verb agreement task~\cite{goldberg2019assessing,clark2019does}, hence it is able to capture long range semantic dependency. Besides, since it is pre-trained with next sentence prediction objective, it remembers context across multiple sentences. Also since 95\% of the sections have less than 512 words, we use the pre-trained BERT model with maximum number of input tokens, to directly encode the sections, hence {\em collapsing} the hierarchy starting from words upto sections in the encoding process. We collapsed the hierarchy up to sections but do not extend further due to several drawbacks in the BERT model discovered through probing~\cite{liu2019linguistic,si2019does} which state that very long sequences spanning multiple sentences leads to incorrect comprehension. We take the final hidden state $h$ of the first token \textsc{[CLS]} as the aggregate representation of the section i.e. $\mathbf{Sc}_x$. (See Figure~\ref{fig:overall_archi}).
 

\subsection{Conditional summarizer}
We next generate the page representation ($\mathbf{D}_{p_i}$, see Figure~\ref{fig:overall_archi}) from the sequence of sections using attention based bidirectional GRU encoder as discussed in the previous section. The inputs $i_x$ in equations (\ref{equ:equ_1}) to (\ref{equ:equ_6}) correspond to the section representations obtained from the BERT models. The $\alpha_x$ correspond to the attention weights, the final output $o_n$ is the page representation $\mathbf{D}_{p_i}$ (see Figure~\ref{fig:overall_archi}).
To aggregate the sequences at section level $Sc_x, x\in[1, n]$, we use the bidirectional GRU~\cite{chung2014empirical} which provides representations of the document text by summarizing information from both directions. We concatenate the forward and backward hidden states $h_x$ and feed the hidden states to the self attention module~\cite{lin2017structured,bahdanau2014neural}.

\footnotesize
\begin{align}
    \overrightarrow{h}_x &= \overrightarrow{GRU}(i_x),\, x \in [1,L] \label{equ:equ_1}\\
    \overleftarrow{h}_x &= \overleftarrow{GRU}(i_x),\, x \in [L,1] \\
    h_x &= [\overrightarrow{h}_x,\overleftarrow{h}_x] \\
    u_x &= \sigma(W_i h_x + b_i) \\
    \alpha_x &= \frac{exp(u_x^T u_i)}{\sum_{x=1}^{L}exp(u_x^T u_i)} \\
    o_n &= \sum_x \alpha_x u_x \label{equ:equ_6}
\end{align}
\normalsize


\subsection{Talk page encoder}
 We split the talk page content into sentences and pass each sentence through the Google Universe Sentence Encoder model~\cite{cer2018universal} to obtain a 512 dimensional representation of the sentence. These sentence embeddings are then passed through a global average pooling layer~\cite{lin2013network} followed by a dense layer to get a 200 dimensional final representation of the talk page ($\mathbf{T}_{p_i}$, see Figure~\ref{fig:overall_archi}). The motivation behind this approach is to obtain an aggregate representation of the discourse in talk pages.
 
\subsection{Fine tuned Inception V3}
\citet{shen2019joint} shows the effectiveness of using the visual rendering (screenshot image) of a page to predict the quality of a document in a multimodal setup. To embed the visual rendering we use the Inception V3~\cite{szegedy2016rethinking} model. Similar to the BERT model, to learn better representations, we fine tune the Inception V3 model. We follow the setting proposed in~\cite{shen2019joint} to fine-tune the Inception V3 model. We resize the images from varied high dimensions to a standard low dimension. We then fine tune the Inception v3 model to classify the screenshots of the pages into the 6 quality classes. We do an end-to-end training for fine tuning the Inception V3 model. We flatten the final representation from the last convolution layer, stack the dropout and global average 2D pooling layers, and classify using softmax layer. The hidden flat layer output (2048 dimension) of the fine tuned Inception V3 model is the final representation of the visual rendering of the page. The input to the fine tuned Inception model is the screenshot of the wikipage, and the output is a visual embedding of the page ($\mathbf{I}_{p_i}$, see Figure~\ref{fig:overall_archi}). We integrate this information with the text and talk content of the page and report the improved performance of the final model.

\subsection{Concatenation module}
In this subsection we illustrate the concatenation of the information from various modules representing the features from text and image modalities to assess the quality of a page. We experiment with different modes of concatenation presented in ~\cite{reimers2019sentence} $(u, v)$, $(u, v, \vert u-v\vert)$, $(u, v, \vert u-v\vert, u*v)$, $(u, v, u*v)$, $(\vert u-v\vert, u*v)$, $(\vert u-v\vert)$ ,$(u*v)$, and choose the best performing strategy $(u, v , \vert u-v\vert)$ to concatenate the vectors $u$ and $v$. We experiment with various combinations of modules, i.e., conditional document summarizer (fine tuned BERT encoder with GRU summarizer), talk page encoder and visual rendering of the page (fine tuned Inception V3) and tabulate the best results in Table~\ref{Tab:Result}.

\subsection{Model configurations}

We set the hidden states ($\overrightarrow{h_x}, \overleftarrow{h_x}$) of the GRU units of the section encoder to 100. 
Therefore the final page representation ($\mathbf{D}_{p_i}$) is a 200 dimensional vector.
Since 90\% of the data has number of sections in a page less than 16 respectively, we limit their maximum size of a page to 16 sections. 
For sequences less than the specified length, we left-pad using $\overrightarrow{0}$. We load the pre-trained weights of BERT-base model from TensorflowHub\footnote{\UrlFont{https://tfhub.dev/google/bert\_uncased\_L-12\_H-768\_A-12/1}}.

We use the implementation of Google Universal Sentence Encoder available at TensorFlow Hub\footnote{\UrlFont{https://tfhub.dev/google/universal-sentence-encoder/}}. We use the nltk library\footnote{\UrlFont{https://www.nltk.org/api/nltk.tokenize.html}} to tokenize the pre-processed talk page content into sentences. We train the hierarchical content encoder for 10 epochs with a learning rate of 0.001 and batches of 16 using Adam optimizer~\cite{kingma2014adam}. For fine tuning the BERT models, we use the Adam optimizer with a learning rate of 2e-5. We empirically set the number of training epochs to 4. To fine tune the Inception V3 model, we again use Adam optimizer with a learning rate 1e-4 and train for 20 epochs. We employ the categorical cross-entropy as loss function for all the models and train using batches of size 16. For all the joint models, we set the learning rate to 0.001, batch size to 32, number of epochs for training to 40. For classification, we use dense layers followed by softmax layer. We further utilize dropout probability of 0.5 in the dense layers. Prior fine-tuning of individual units reduces explosive training time, common in end-to-end models

\section{Experiment}
In this section we evaluate {\bf NwQM} against several existing approaches.
\paragraph{Baselines.}
 We provide a brief outline of the competing methods in the following.
\begin{compactitem}
\item \textbf{ORES} \cite{halfaker2019ores} is a machine learning service made available by Wikimedia foundation through a RESTful HTTP interface serving prediction about target articles. It uses handcrafted features along with gradient boosted machine as classifier.

\item \textbf{\textsc{doc2vec}\xspace}~\cite{dang2016quality} proposed the first application of deep neural networks into quality assessment task where they employed distributional representation of documents~\cite{le2014distributed} without using manual features. 

\item \textbf{\bilstm}+ \cite{shen2017hybrid} is a hybrid model, where textual content of the Wikipedia articles are encoded using a \bilstm model. The hidden representation captured by the sequence model is further augmented with handcrafted features and the concatenated feature vector is used for final classification
\item \textbf{H-\lstm}~\cite{zhang2018history} is an edit history based approach where every version of an article is represented by $17$ dimensional handcrafted features. Hence an acticle with $k$ versions will be represented by $k \times 17$ matrix. This $k$ length sequence is passed through a stacked \lstm for final representation used in classification.
\item \textbf{M-\bilstm}~\cite{shen2019joint} proposed a multimodal information fusion approach where embeddings obtained from both article text as well as html rendering of the article webpage is used for final classification. 
\item \textbf{\textsc{DocBert}\xspace}~\cite{adhikari2019docbert} proposed this method to generate document representations using bidirectional transformers~\cite{devlin2018bert}. The primary idea is filtering the representation obtained from the \textsc{CLS} token using a fully connected layer which translates 768 dimensional encoding to class distribution using a softmax layer. This architecture is further fine tuned end-to-end for the respective document classification task.

\item \textbf{\textsc{HAN}\xspace}~\cite{yang2016hierarchical} proposed a hierarchical approach which iteratively constructs a document
vector by coalescing important words into sentence
vectors and subsequently combining important sentences
vectors to obtain the document vectors. A convenient consequence of this approach is that it is suitable for large documents like Wikipedia articles.

\end{compactitem}

\paragraph{Result}
We evaluate \textbf{NwQM} against existing solutions for automatic quality assessment and tabulate the obtained results in Table~\ref{Tab:Result}. Since our classes are roughly balanced, we opt to report accuracy as the metric for evaluation. Some of the approaches that we compare our model with are \textsc{ores}, \textsc{doc2vec}, \textsc{bilstm+}, M-\textsc{lstm}, H-\textsc{lstm}. We achieve an improvement of $8\%$ compared to the best performing baseline. This we believe is a considerable leap for a 6 class classification task. 

We also compare our model against novel document classification approaches, i.e., \textsc{docbert} and \textsc{Han} because inherently Wikiepedia quality assessment problem is closely related to document categorization. To this purpose, we compare the document classification approaches with textual content of the Wikipedia main article as well as concatenated version of the main article and talk pages denoted as \textsc{docbert}-wT, \textsc{han}-wT respectively. We obtain at most $5\%$ improvement against the existing approaches. Note that apart from our model, including the talk page meta data in the document classification models also considerably enhances their respective performances. 

\textbf{NwQM} is constituted of concatenated representation from article text, talk and image and therefore it is important to look at how individual components perform independently. We evaluate \textbf{NwQM} without signals from image (\textbf{NwQM}-w/oI), without talk (\textbf{NwQM}-w/oT) and with solely the main article text, i.e., without any secondary and tertiary signals from image and metadata (\textbf{NwQM}-w/oTI). Our experiment show that the combined approach (\textbf{NwQM}) obtains the best result. Without talk and image we land in a drop of accuracy of $1.3\%$ and $5\%$ respectively validating our hypothesis that extracting signals from external sources can serve fruitful in this task. We also compare against representation from talk pages and fine tuned image embeddings individually for the sake of completeness and our results show significant drop in accuracy compared to the combined approach.

\begin{table}[t]\footnotesize\setlength{\tabcolsep}{12pt}
    \centering
    \begin{tabular}{c c}
    \toprule
    Model &  Accuracy  \\
    \midrule
     \textsc{ORES}~\cite{halfaker2019ores} & 43.21  \\
     \textsc{doc2vec} wRF~\cite{dang2016quality} & 44.01  \\
     \textsc{doc2vec} wLR~\cite{dang2016quality} & 49.33  \\
     \textsc{bilstm}+~\cite{shen2017hybrid} & 54.5  \\
    H-\textsc{\lstm}~\cite{zhang2018history} & 53.05  \\
    \rowcolor{blue!20}M-\bilstm~\cite{shen2019joint} & 58.47  \\
    \textsc{docbert}~\cite{adhikari2019docbert} & 57.66 \\
    \rowcolor{red!20}\textsc{docbert}-wT~\cite{adhikari2019docbert} & 59.87 \\
    \textsc{HAN}~\cite{yang2016hierarchical} & 56.35 \\
    \textsc{HAN}-wT~\cite{yang2016hierarchical}& 57.48 \\
    \rowcolor{green!20}\textsc{NwQM} & 63.23 \\
    \textsc{NwQM-w/oI} & 59.95 \\
    \textsc{NwQM-w/oT} & 62.37 \\
    \textsc{NwQM-w/oTI} & 59.10 \\
    \textsc{Talk} & 37.95 \\
     \textsc{Inception V3} & 52.96 \\
    \bottomrule
\end{tabular}
		\caption{\label{Tab:Result} Results obtained from different models. The best result is highlighted in \textcolor{green}{green}, the best result among document classification models is highlighted in \textcolor{red}{red} and the best result among the state-of-the-art quality assessment models is highlighted in \textcolor{blue}{blue}.}
	\end{table}
\vspace{-2mm}
\begin{figure}[ht]
\centering
  \begin{tabular}{l l}
\centering
    \includegraphics[width=0.22\textwidth,height=0.2\textwidth]{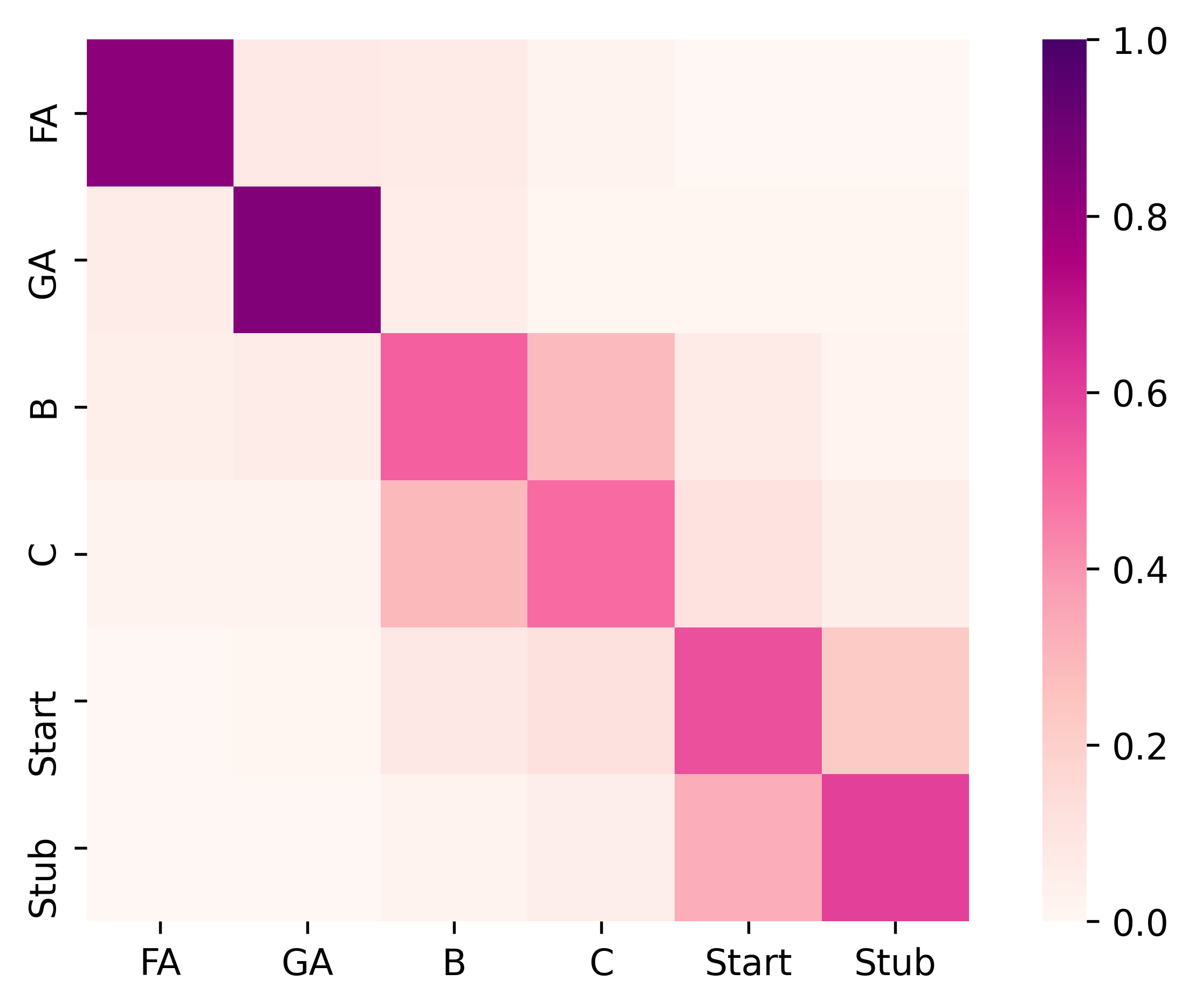} & \includegraphics[width=0.22\textwidth,height=0.2\textwidth]{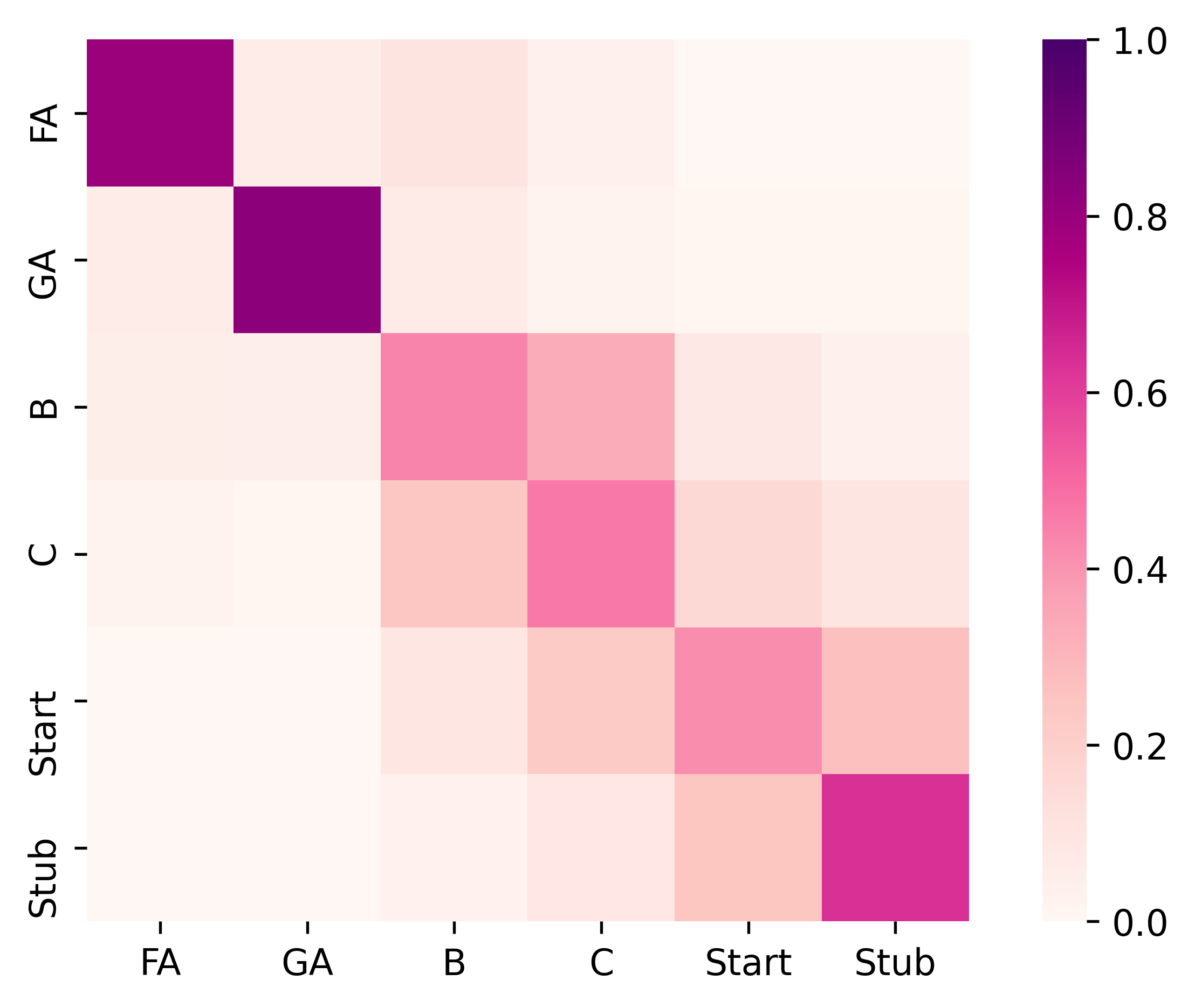} \\ 
  \end{tabular}
  \caption{Left panel shows confusion matrix obtained by \textbf{NwQM}; Right panel shows confusion matrix obtained by M-\textsc{bilstm} (Best viewed in color) .}
  \label{fig:confusion_matrix}
\end{figure}

\begin{figure}[!ht]
    \centering
    \includegraphics[width=1.0\columnwidth,height=0.23\textwidth]{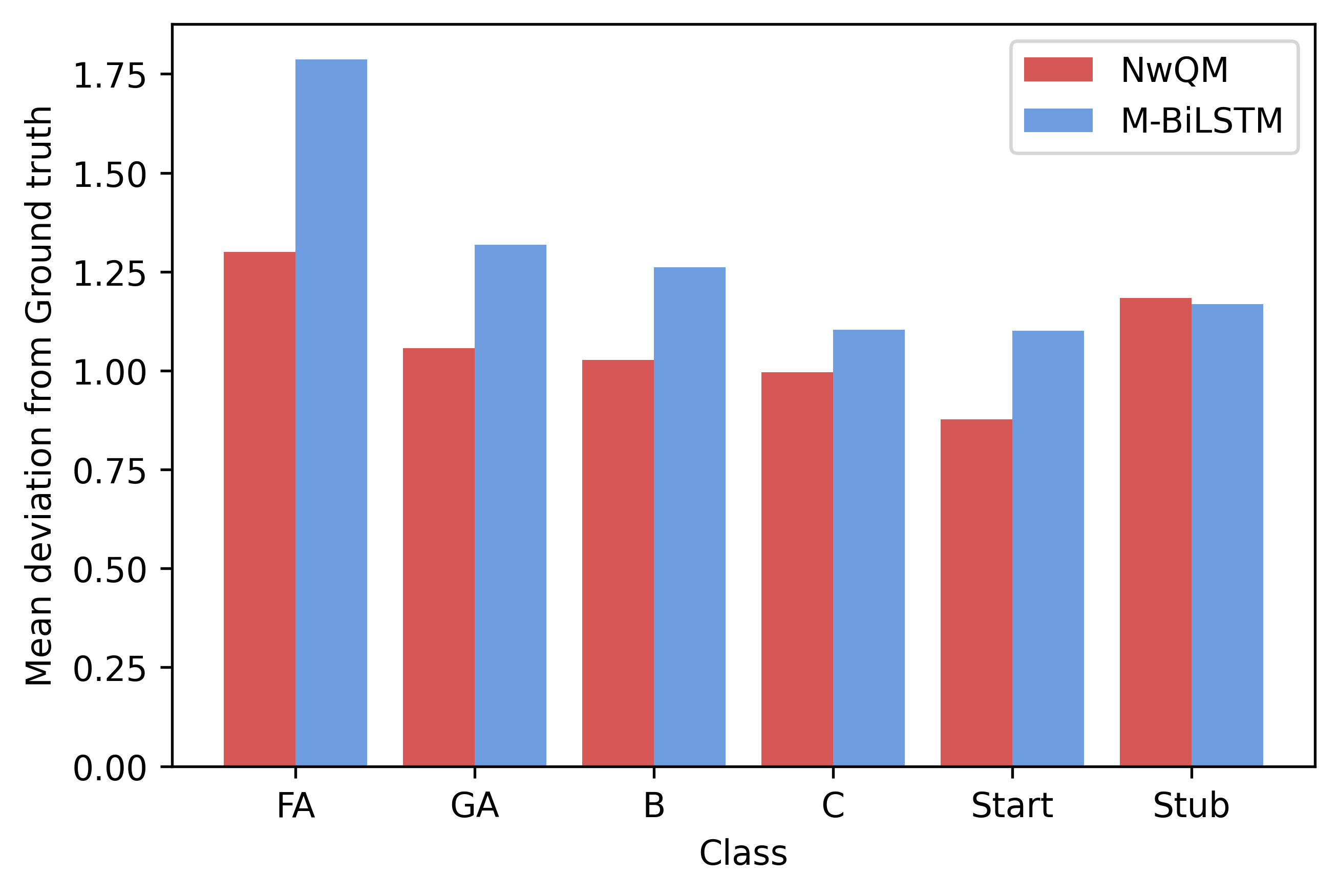}
    \caption{Mean absolute distance of the misclassifications from the true labels for individual classes. (Best viewed in color)}
    \label{fig:success_stats}
\end{figure}

\begin{figure*}[!ht]
\centering
  \begin{tabular}{l l}
\centering
    \includegraphics[width=0.45\textwidth,height=0.175\textwidth]{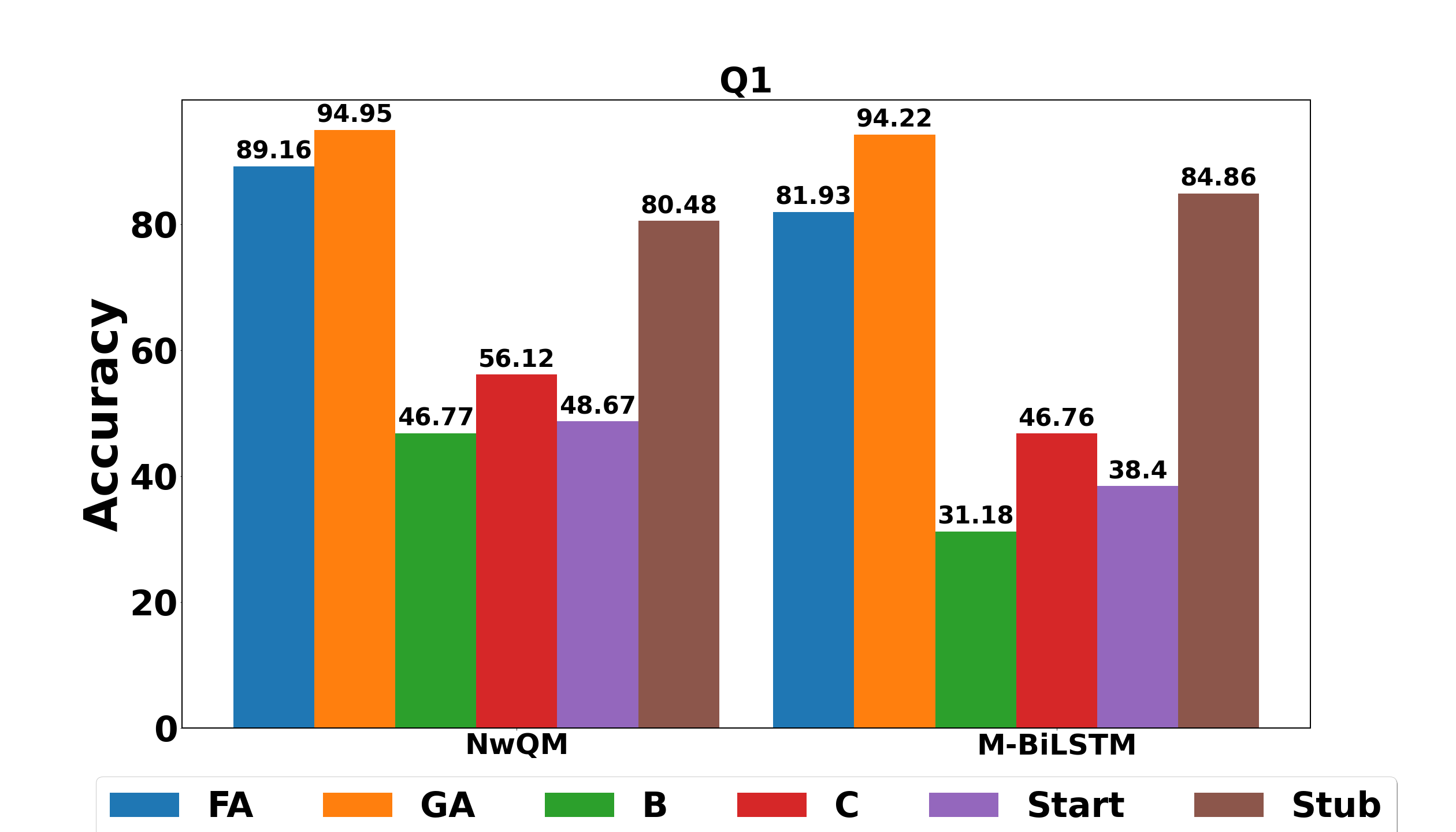} & \includegraphics[width=0.45\textwidth,height=0.175\textwidth]{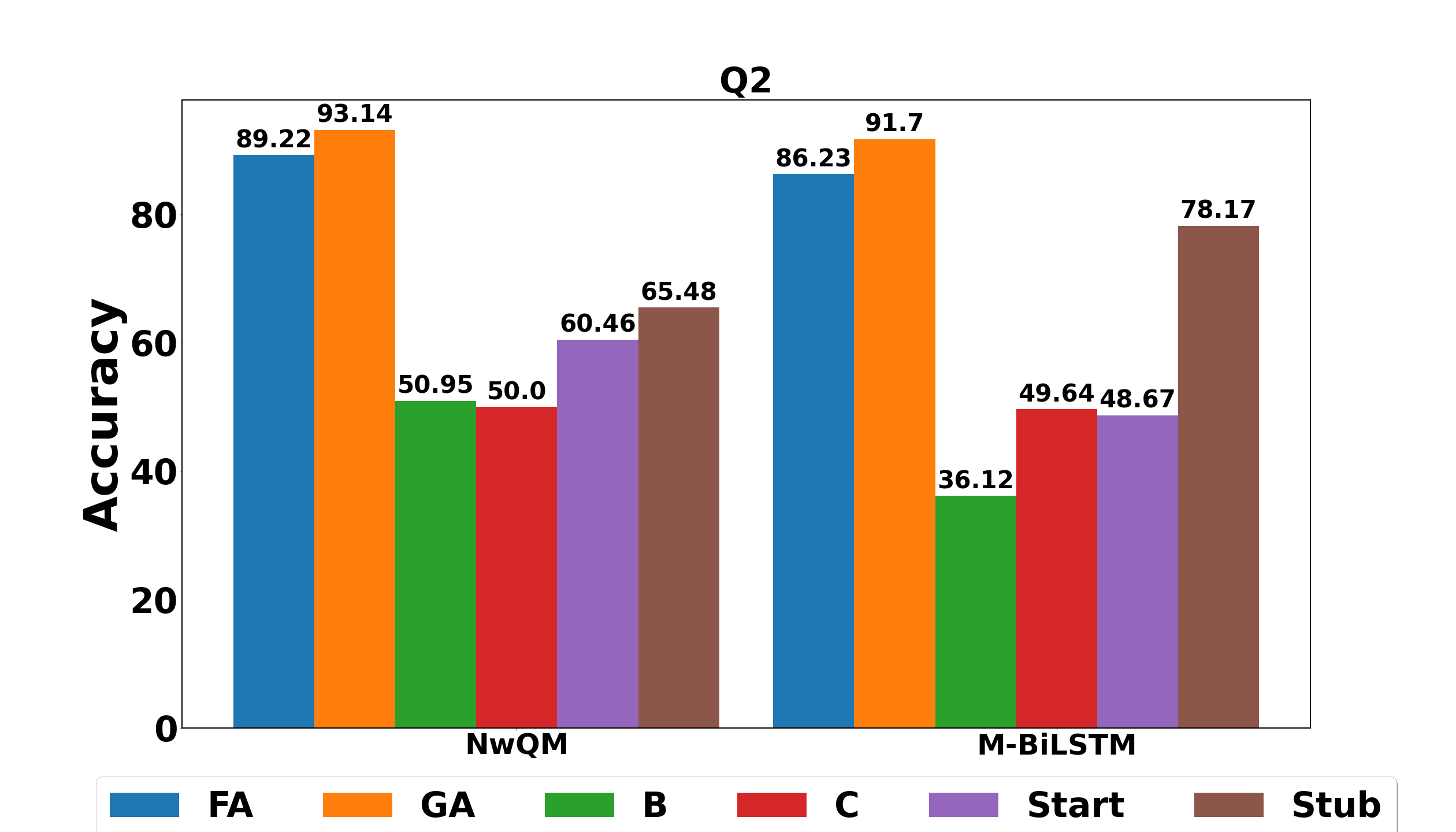} \\ \includegraphics[width=0.45\textwidth,height=0.175\textwidth]{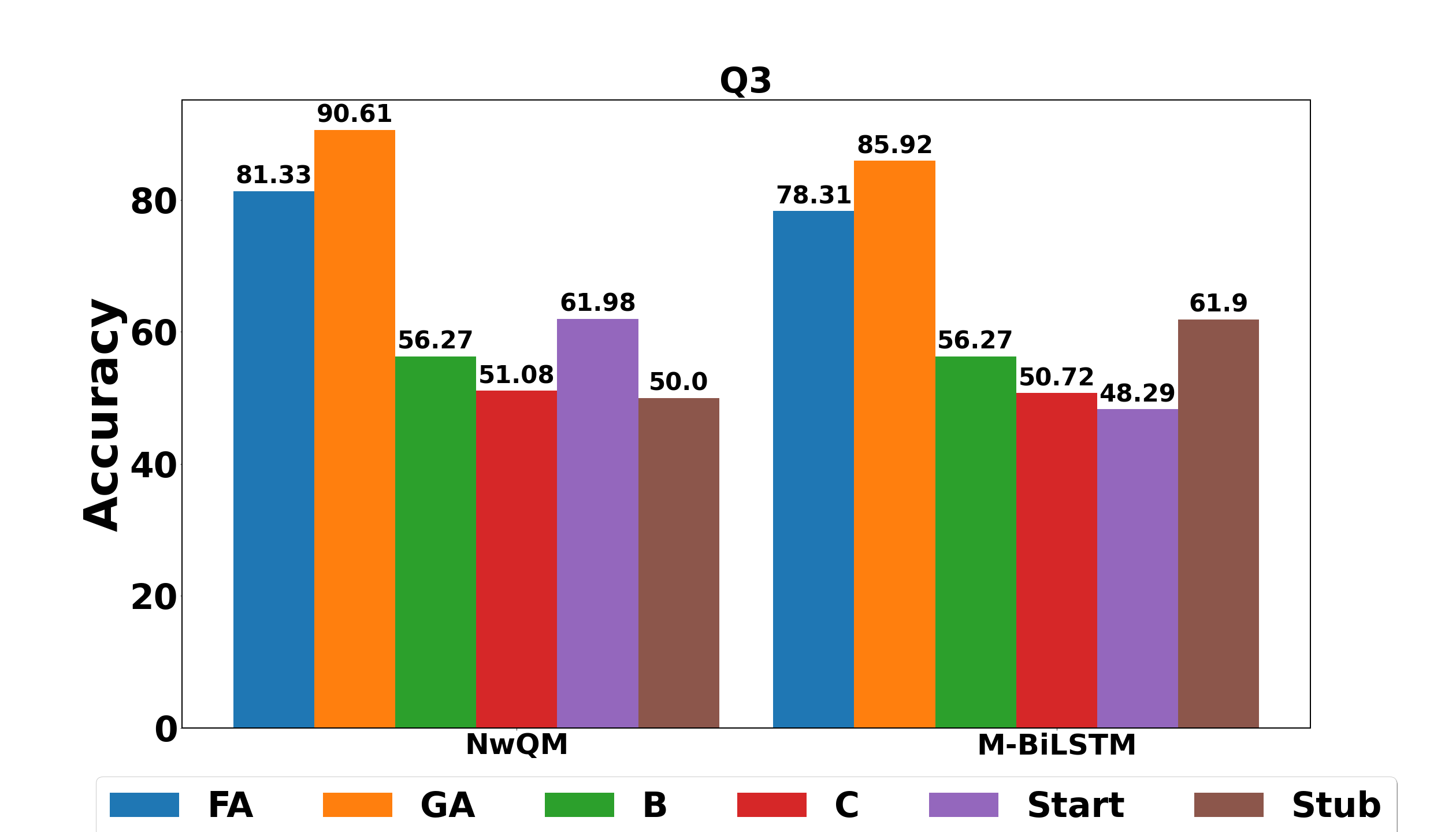} & \includegraphics[width=0.45\textwidth,height=0.175\textwidth]{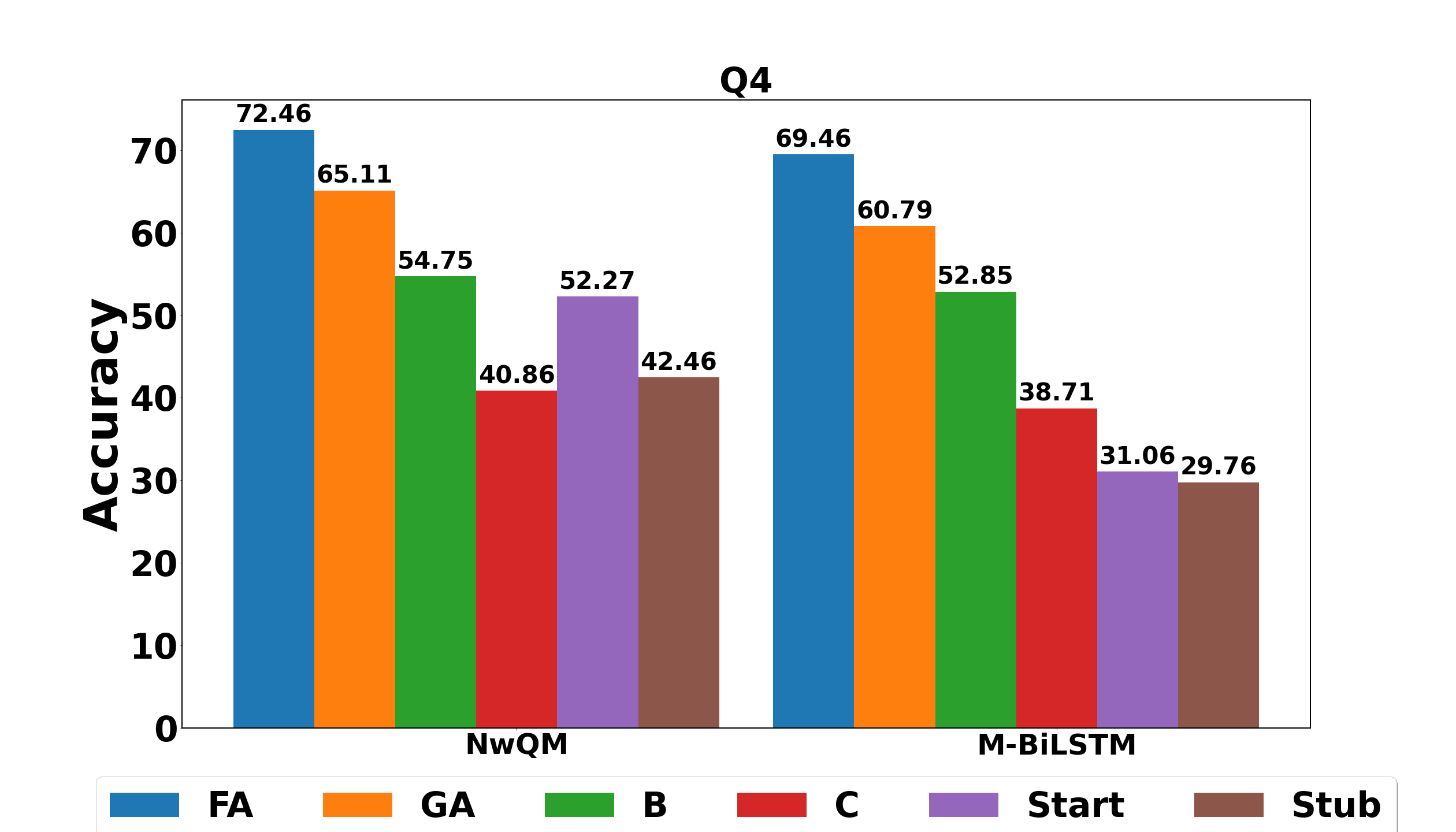}  \\
  \end{tabular}
  \caption{Accuracy per class for test data segregated into quartiles $Q_1,Q_2,Q_3,Q_4$ with respect to main article text length with $Q_1$ smallest and $Q_4$ largest. (Best viewed in color)}
  \label{fig:quality_result3a}
\end{figure*}

\begin{figure*}[!ht]
\centering
  \begin{tabular}{l l}
\centering
    \includegraphics[width=0.45\textwidth,height=0.175\textwidth]{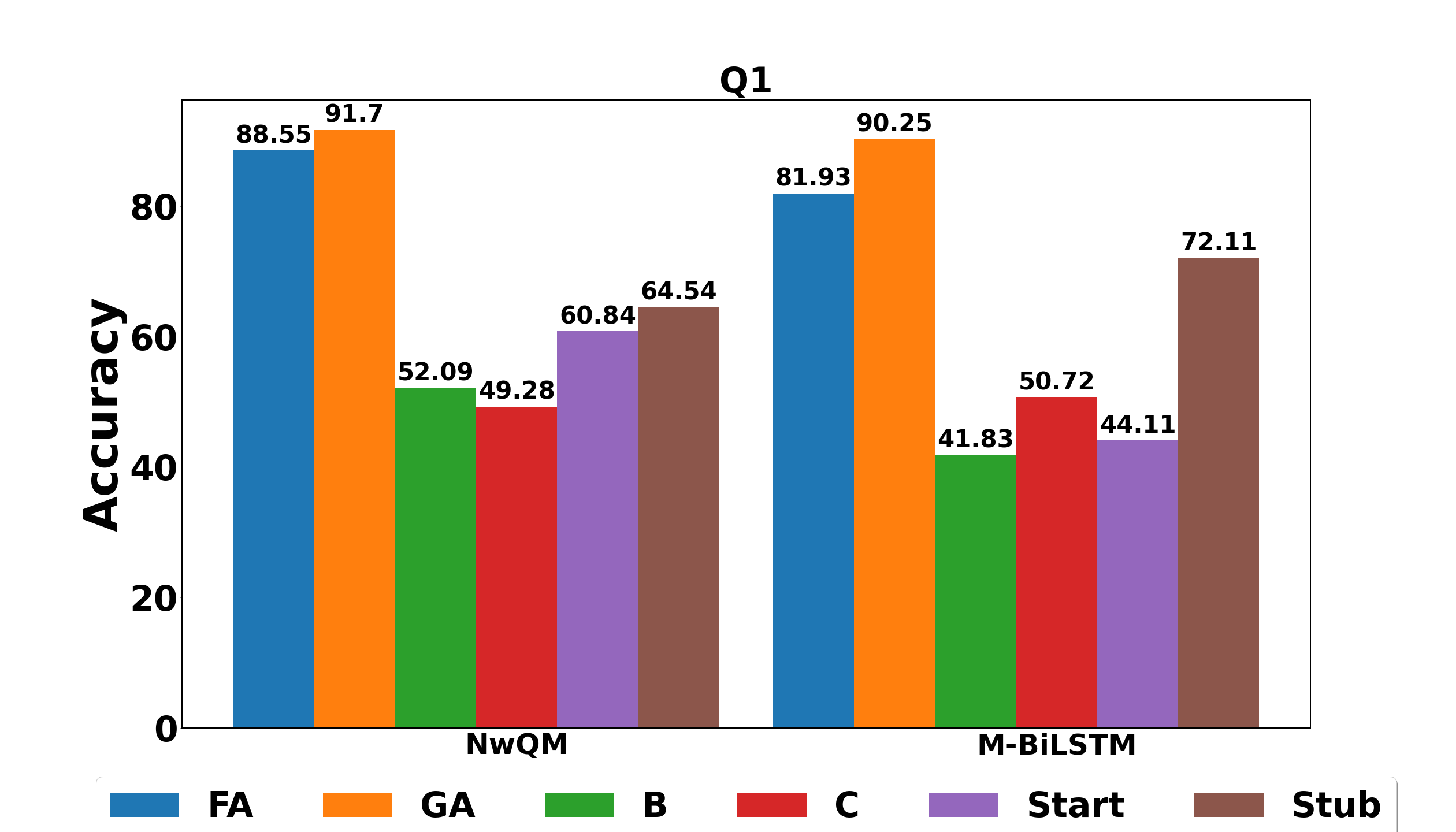} & \includegraphics[width=0.45\textwidth,height=0.175\textwidth]{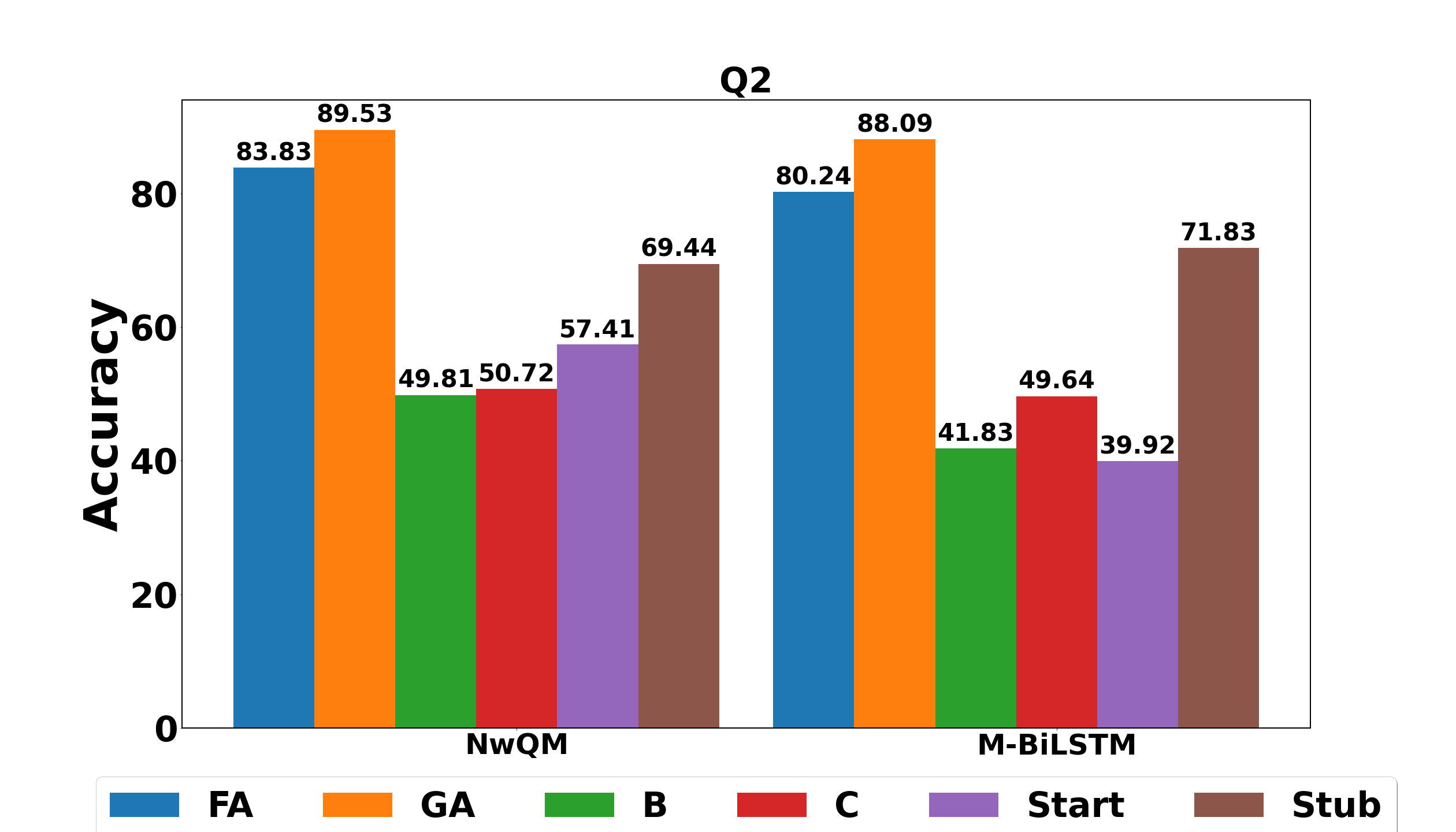} \\ \includegraphics[width=0.45\textwidth,height=0.175\textwidth]{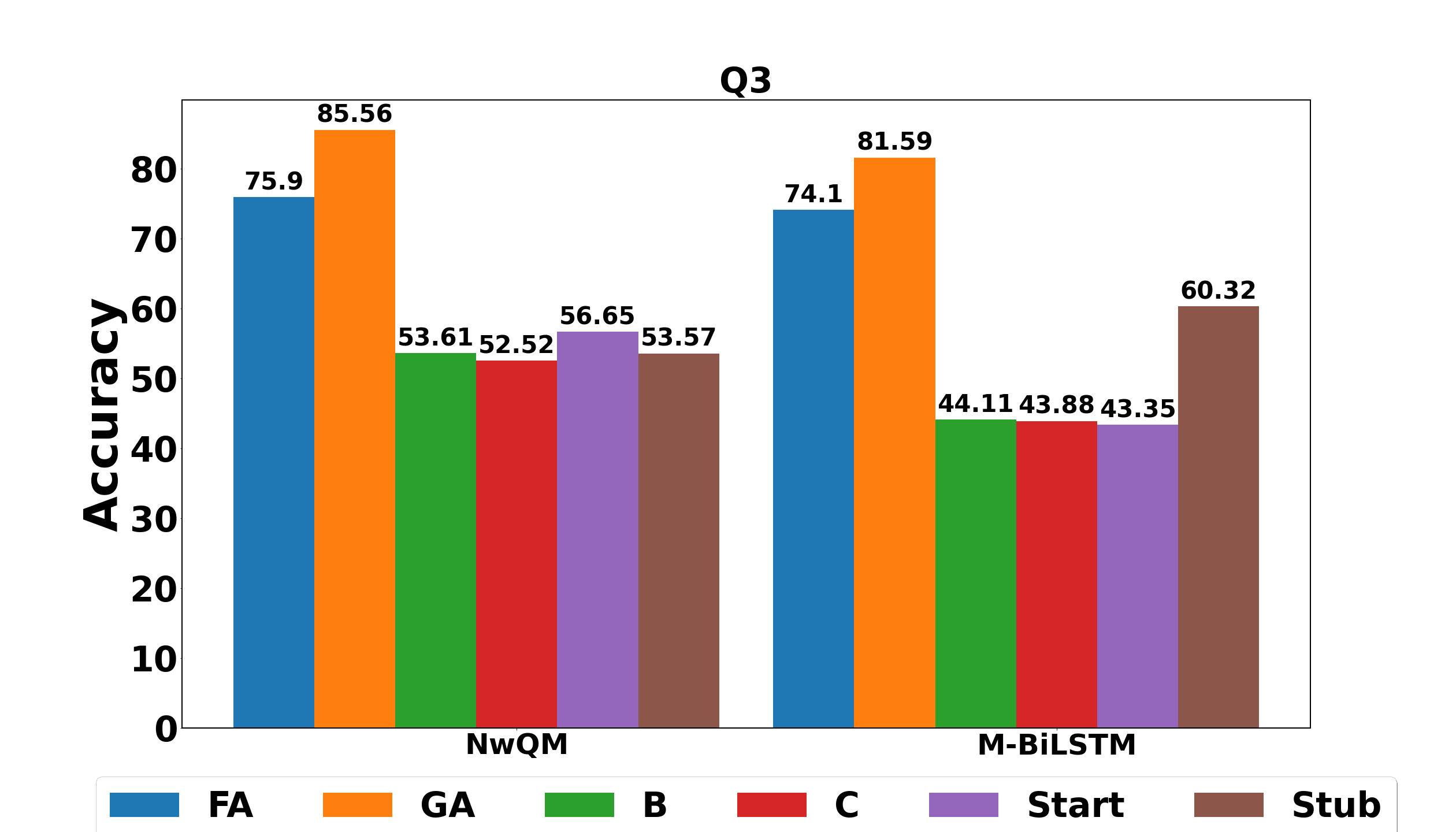} & \includegraphics[width=0.45\textwidth,height=0.175\textwidth]{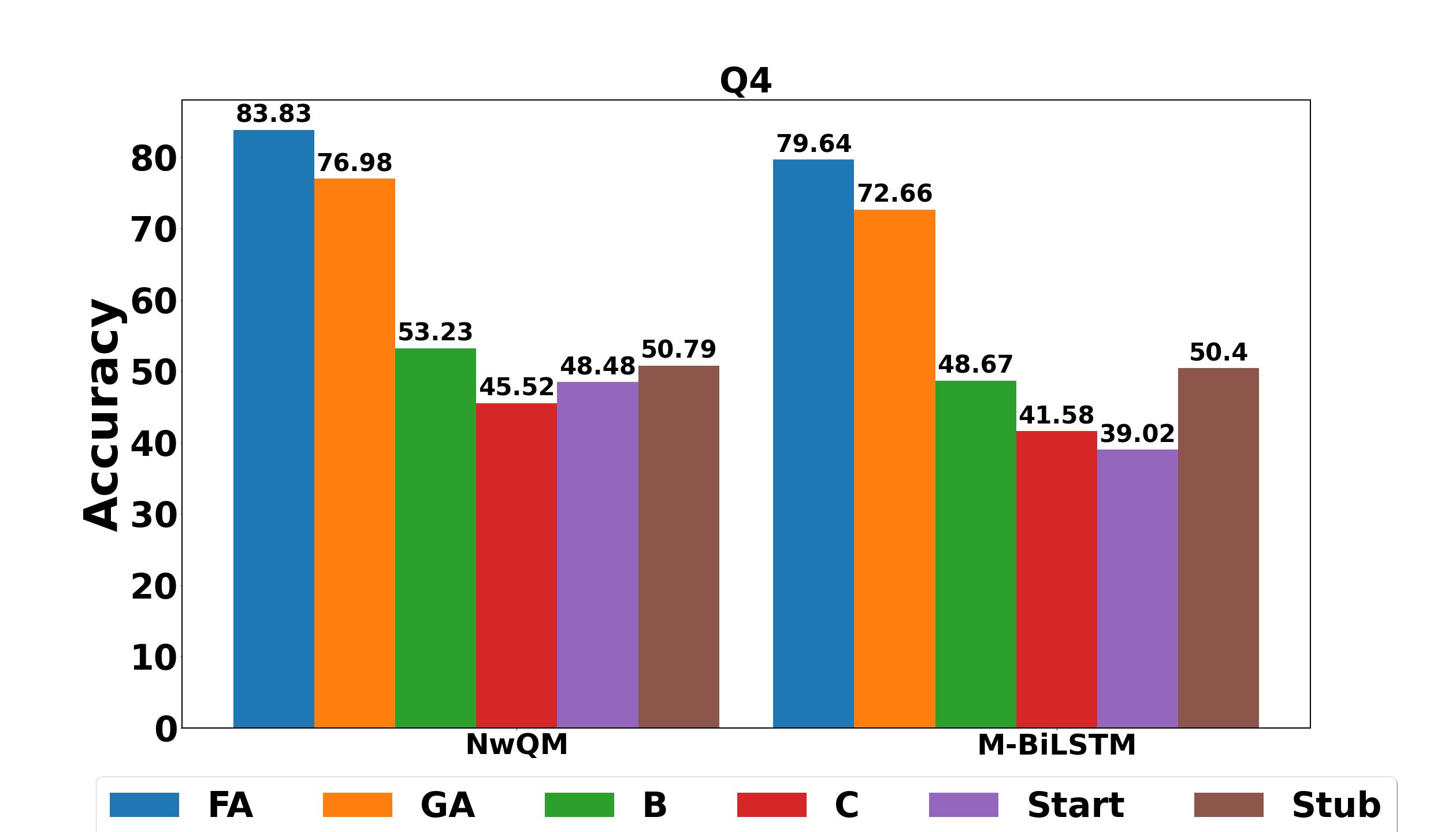}  \\
  \end{tabular}
  \caption{Accuracy per class for test data segregated into quartiles $Q_1,Q_2,Q_3,Q_4$ with respect to article talk length with $Q_1$ smallest and $Q_4$ largest. (Best viewed in color)}
  \label{fig:quality_result3b}
\end{figure*}

\section{Discussion and qualitative analysis}
In this section we perform a deep dive into the predictions obtained by our model and we contrast it with those from closest known competitor M-\textsc{bilstm}~\cite{shen2019joint} for quality assessment. 

\noindent\textbf{Confusion matrix}: We first tabulate the confusion matrix obtained by \textbf{NwQM} on test data in Figure~\ref{fig:confusion_matrix}. Results show that most of the misclassifactions made by \textbf{NwQM} are on average between very similar classes. This is due to the inherent ordinal nature of the classes in this dataset. More specifically, there is an increasing degree of quality in pages from Stub to FA, though the gradient may not be smooth. Thus FA, GA articles have overlapping guidelines which  is very different from other pages. Likewise B, C and Start, Stub have mutually overlapping guidelines. \textbf{NwQM} can successfully recover this structure, hence misclassifications have occurred between closer classes. However, compared to M-\textsc{bilstm} (see Figure~\ref{fig:confusion_matrix}) \textbf{NwQM} can capture class specific features significantly better thus showing lower mistakes for certain datapoints especially in case of Start, B and C classes.

\noindent\textbf{Distance of wrong predictions from ground truth}: In cases of wrong predictions, we calculate how far individual models are from ground truth labels. We transform each class into integers with Stub transformed to $0$ and FA transformed to $5$. We then find the mean absolute distance of the incorrect predictions from the true labels for each individual class. The results are tabulated in Figure~\ref{fig:success_stats} where we observe that for \textbf{NwQM} the absolute distance is consistently lower. The results are statistically significant with a $p$-value of 1e-5 using Stuart-Maxwell test for multiclass classification~\cite{sun2008generalized}.

\noindent\textbf{Effect of article length}: We further ascertain how the main text and talk page length of an plays a role in prediction, for respective models. Articles with detailed coverage of topics and multiple discussion threads may not be nominated among high quality articles. This is because even if the coverage is diverse, the language of the article may be biased, convoluted, there could be unverified claims or disaccord among editors. Deep learning models such as BERT have often been shown to rely on surface forms~\cite{ettinger2020bert} as shotcuts for classification instead of semantic understanding. We investigate whether \textbf{NwQM} is relying spurious signals like article length in final classification. For example, article {\em Kauri Gum} (GA) and {\em Guar Gum} (Start) have very similar coverage, but are distant in terms of the quality criteria. Similarly, article {\em Frog cake} (GA) and {\em Sugar} (GA) have very different coverage but the same quality tag. For further investigation we rank predictions with respect to article and talk page length and divide our predictions into $4$ quantiles. We investigate the accuracy of \textbf{NwQM} in the individual quantiles starting from the smallest quantile, i.e., {\em $Q_1$} to the largest, i.e., {\em $Q_4$}. Accuracy scores for test set ordered by main article text are illustrated in Figure~\ref{fig:quality_result3a} and those ordered based on talk pages are shown in Figure~\ref{fig:quality_result3b}. Results indicate that \textbf{NwQM} is not biased toward length and irrespective of the quantiles we obtain improved prediction accuracy almost always compared to the closest baseline.

\begin{figure}[!ht]
\centering
  \begin{tabular}{l}
\centering
    \includegraphics[width=0.45\textwidth,height=0.2\textwidth]{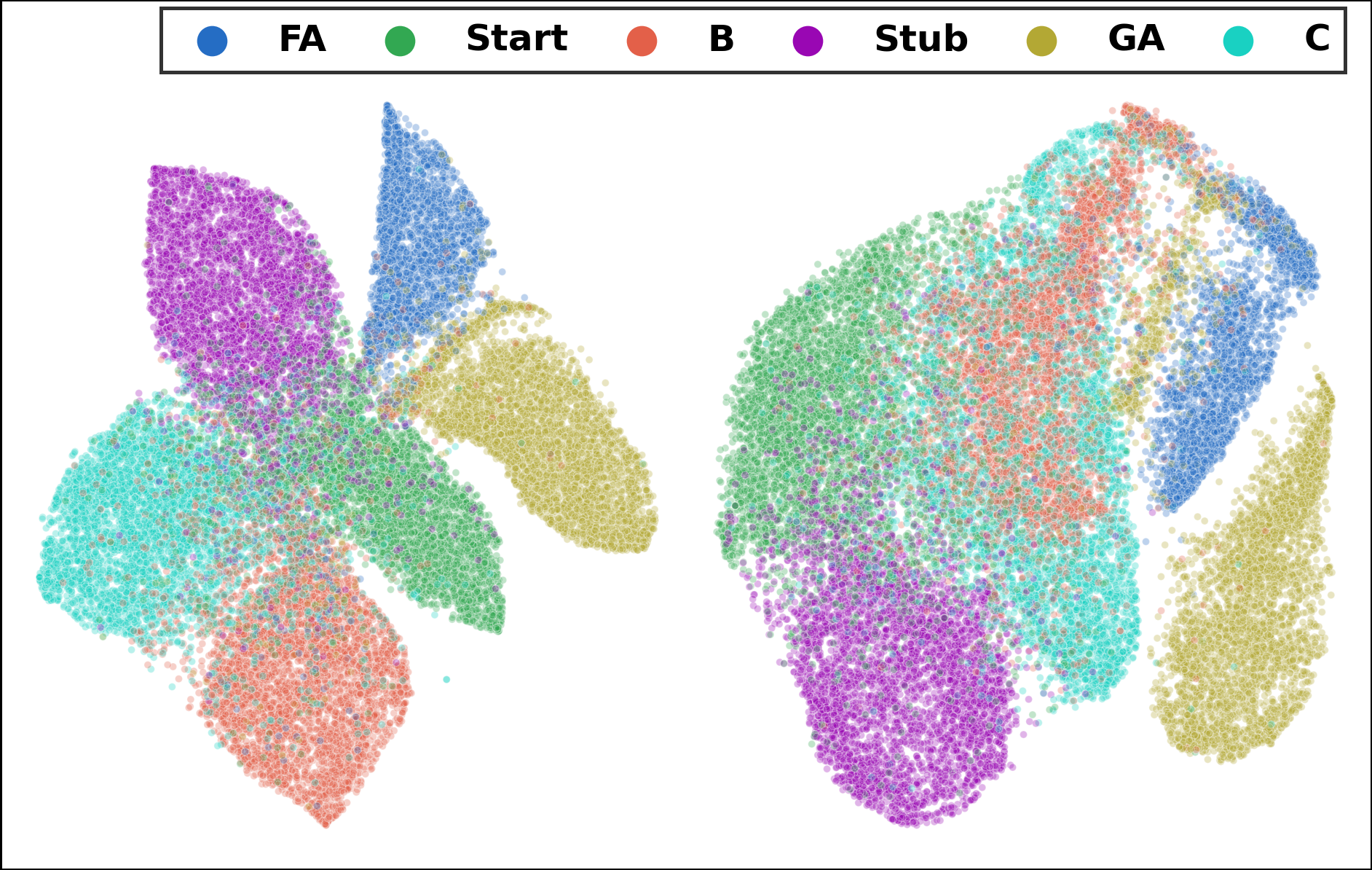}   
  \end{tabular}
  \caption{Left panel shows t-SNE visualization for \textbf{NwQM}; right panel shows visualization for M-\textsc{bistm}. (Best viewed in color)}
  \vspace{-3mm}
  \label{fig:quality_result2}
\end{figure}

\noindent\textbf{t-SNE plots}: Finally, we visualize the representations obtained by \textbf{NwQM} and M-\textsc{bilstm} using t-SNE~\cite{maaten2008visualizing} scatter plots (see Figure~\ref{fig:quality_result2}). The degree of separation obtained by \textbf{NwQM} is much better which translates to the improved accuracy. 

\begin{figure}[!ht]
\centering
  \begin{tabular}{l}
\centering
    \includegraphics[width=0.5\textwidth,height=0.2\textwidth]{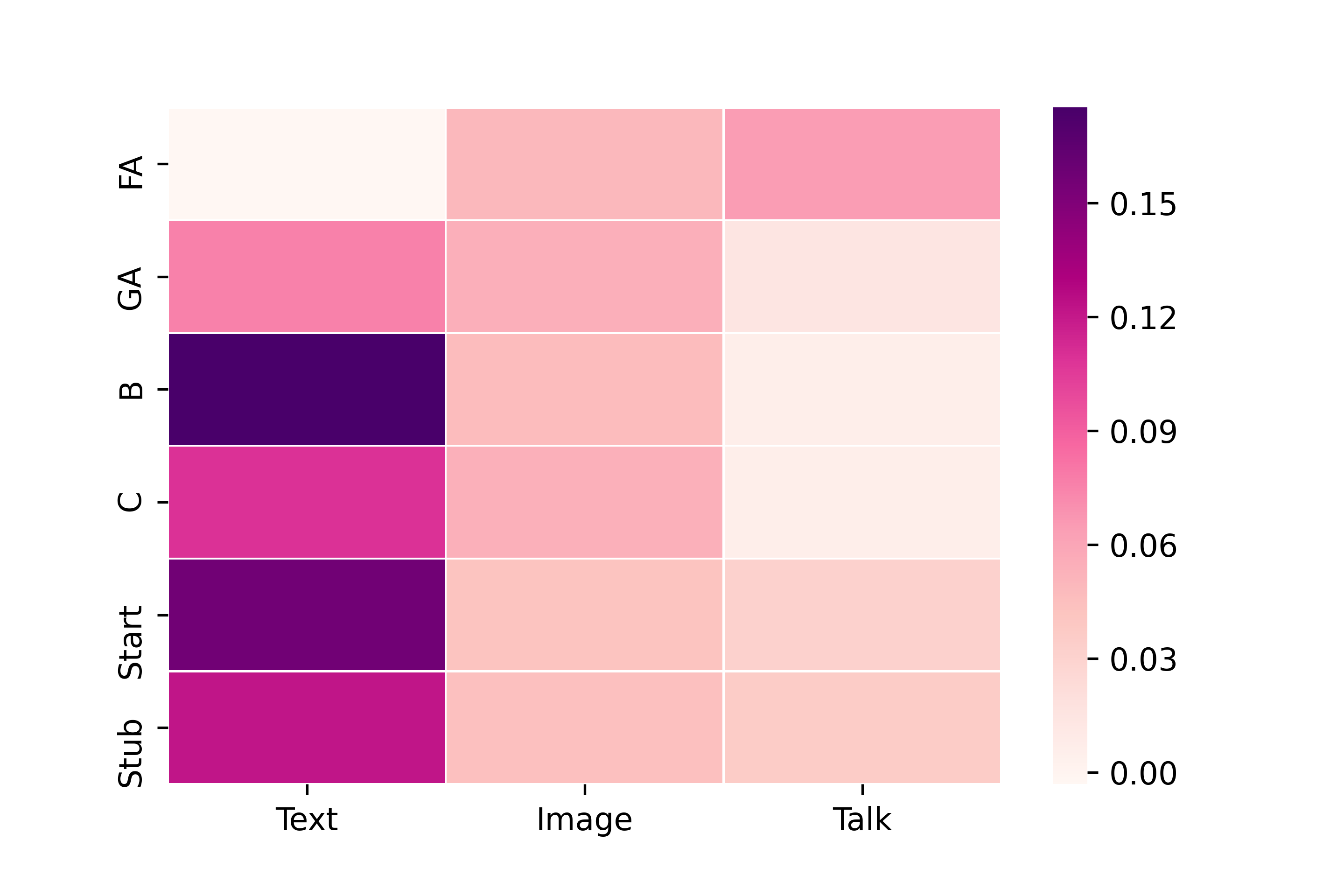}  
  \end{tabular}
  \caption{ (Best viewed in color)}
  \vspace{-3mm}
  \label{fig:heat_map}
\end{figure}

\noindent\textbf{Interpreting the effect of the different modalities}: One of the critical issues in deep learning models is interpretability. In order to ascertain whether different sources of signals are indeed contributing toward the final prediction task, we leverage the model agnostic evaluation tool LIME~\cite{ribeiro2016should}. We generate representations for text, talk and images, i.e., $D_{p_i},T_{p_i},I_{p_i}$ respectively (see Figure~\ref{fig:overall_archi}) from our learned model. These embeddings are passed through classification layers comprising dense layer and softmax layer for prediction. This network takes concatenated input of $D_{p_i},T_{p_i},I_{p_i}$ and outputs the classification probabilities for each test instance. We evaluate this black box neural network using LIME. For data points on the test set we identify top 500 features contributing to the outcome of the highest class probability. We further calculate the average contribution from each modality toward the respective classes. More specifically for every test page instance, we identify the top 500 contributing features as per LIME. Each feature can be contributed by any one of the three modalities. We compute the mean of features scores per modality in the top 500 for a page and then aggregate that over all pages. The results are tabulate in Figure~\ref{fig:heat_map}. Our results show that signals from $D_{p_i}$ play most important overall role in prediction of quality. $I_{p_i}$ play almost equal role in prediction of all the classes. Interestingly, for high quality pages, $T_{p_i}$, i.e., the talk page information contributes in classification higher than text and image information. Talk also contributes for low quality pages such as {\em Start, Stub}. We speculate that since high quality pages have larger discussion archives and low quality pages have very low discussion threads $T_{p_i}$ plays significant role in distinguishing these classes. We stress that these interpretability results are the prime insights that these paper neatly establishes. 

\section{Conclusion}
\vspace{-2mm}
In this paper we proposed a novel multimodal deep learning based model \textbf{NwQM} for quality assessment of English Wikipedia articles. Our model combines signals from article text, meta pages and image rendering to construct an improved document representation.  We evaluate it against several existing approaches and obtain at most $8\%$ improvement compared to the state-of-the-art method. For a 6 class classification this leap in accuracy is notable. We also perform extensive investigation of the different components of our model to understand their individual utility. We perform in-depth qualitative analysis of the obtained predictions and contrast them with the closest baseline.

To the best of our knowledge this is the first work which combines several aspects of information available for Wikipedia articles and, in particular, the talk page dynamics toward quality assessment. We also showcase the utility of fine tuned bidirectional transformers toward document classification especially when combined with niche platform specific signals. We believe our work opens up the necessity of further investigation pertaining to careful information fusion techniques for downstream tasks.  

\bibliography{emnlp2020}
\bibliographystyle{acl_natbib}
\end{document}